\RequirePackage{fix-cm}
\documentclass[twocolumn]{svjour3}          
\smartqed  
\usepackage{graphicx}%
\usepackage{latexsym}
\usepackage{amsmath}
\usepackage{amssymb}
\usepackage{subfig}
\usepackage{graphicx}
\usepackage{dcolumn}
\usepackage{bm}
\usepackage{color} 	
\usepackage{textcomp}
\newcommand{\la}{\lambda}
\newcommand{\ka}{\kappa}

\newcommand\mcn{\mathcal{N}}
\providecommand\bvarpi{\pmb{\varPi}}
\providecommand\bcdot{\pmb{\cdot}}
\newcommand\bi{\pmb{I}}
\newcommand\bD{\pmb{D}}
\newcommand{\dx}{\frac{{\rm d} }{{\rm d} x}}
\definecolor{orange}{rgb}{1,0.5,0}
\definecolor{pinegreen}{rgb}{0,1,0.95}
\definecolor{darkgreen}{rgb}{0,0.54,0}
\definecolor{gray}{rgb}{0.3, 0.3, 0.3}
\journalname{Shock Waves}
\begin{document}

\title{Reinterpreting Shock Wave Structure Predictions using the Navier-Stokes Equations
}

\titlerunning{Reinterpreting Shock Wave Structures}        

\author{M.~H.~Lakshminarayana Reddy \and
S.~Kokou~Dadzie 
}

\authorrunning{M.~H.~L.~Reddy \and S.~K.~Dadzie} 

\institute{M.~H.~Lakshminarayana Reddy \at
                 School of Engineering and Physical Sciences\\
                 Heriot-Watt University\\
                 Edinburgh EH14 4AS\\
                 Scotland, UK\\
              \email{l.mh@hw.ac.uk}           
           \and
           S.~Kokou~Dadzie \at
                    School of Engineering and Physical Sciences\\
                    Heriot-Watt University\\
                    Edinburgh EH14 4AS\\
                    Scotland, UK\\
              \email{k.dadzie@hw.ac.uk}
}

\date{Received: date / Accepted: date}

\maketitle

\begin{abstract}
Classical Navier-Stokes equations fail to predict shock wave profiles accurately.
In this paper, the Navier-Stokes system is fully transformed using a velocity variable transformation. The transformed equations termed the recast Navier-Stokes equations display physics not initially included in the classical form of the equations. We then analyse the stationary shock structure problem in a monatomic gas by solving both the classical and the recast Navier-Stokes equations numerically using a finite difference global solution (FDGS) scheme. The numerical results are presented for different upstream Mach numbers ranging from supersonic to hypersonic flows. We found that the recast Navier-Stokes equations show better agreements with the experimentally measured density and reciprocal shock thickness profiles.
\end{abstract}
\section{Introduction}
\label{intro}
One of the best-known examples of a simple and highly non-equilibrium compressible flow phenomenon is that of a normal shock wave. A normal shock wave is a disturbance propagating between a supersonic fluid and a subsonic fluid, characterized by a sharp change in its fluid properties. In other words, one can treat the shock wave as an interface of finite thickness between two different equilibrium states of a gas \cite{CF1948,Grad1952,Bird1994,Reese1995,ReddyAlam2015}. Shock waves arise at explosions, detonations, supersonic movements of bodies, and so on. The shock structure problem has been studied extensively in the middle of the $20^{th}$ century using theoretical, numerical and experimental techniques. It now serves as a standard benchmark problem for testing the capability (validity) and accuracy of different hydrodynamics and extended hydrodynamic fluid flow models \cite{Greenshields2007,Reddy2016,ReddyAlam2020}. A few advantages of a shock structure problem making it attractive for numerical simulations are: (i) it is one-dimensional and steady state; (ii) the upstream and downstream boundary conditions are clearly specified by the Rankine-Hugoniot conditions; (iii) all gradients of hydrodynamic field variables vanish far upstream and downstream of the shock; and (iv) solid boundaries are absent \cite{LeVeque2002}.

The principal parameter used to classify the non-equilibrium state of a rarefied flow is the Knudsen number, $\rm{Kn}$. It is defined as the ratio of the mean free path of the gas molecules to the characteristic length of the flow system. In the shock structure problem, $\rm{Kn}$ is related to the shock thickness \cite{Greenshields2007}. Within a shock layer, physical properties of the gas change very fast over a distance of a few mean free paths which makes the Knudsen number large. Typical values of the Knudsen number for a flow within a shock layer fall between $\approx 0.2$ and $\approx 0.3$ \cite{Greenshields2007}. These are beyond the classical continuum-$\rm{Kn}$ regime and fall into the so-called `intermediate-$\rm{Kn}$' regime ($0.01 \lesssim \rm{Kn} \lesssim 1$). Hence shock structures are not well captured by standard fluid dynamic equations. In particular, shock structure predictions from standard Navier-Stokes equations have shown some agreement with the experimental data at low Mach numbers ${\it{M}}_{\rm 1} < 1.5$ but clearly failed above Mach number of $2$ \cite{Alsmeyer1976}. Deriving appropriate continuum models that can predict these data is therefore still an active research topic \cite{Paolucci2018}.

In this article, a method is used to reinterpret the original Navier-Stokes equations and its prediction of the experimental data. A change of velocity variable is used to transform the equations into physically different equations before they are solved to compare with the experimental data.

The paper is arranged as follows: in section \ref{2} we briefly present the classical hydrodynamic equations of fluid flows along with the new strategy to obtain a new continuum hydrodynamic model, namely, the recast Navier-Stokes equations. In the following section \ref{3}, both hydrodynamic models are reduced to a one-dimensional stationary shock structure problem and then solved numerically using a finite difference global solution scheme (FDGS). Predictions of shock structures by both models are presented and compared with existing experimental and direct simulation Monte Carlo (DSMC) data in section \ref{4}. At the end, conclusions are presented.

\section{\label{2} The classical and the recast Navier-Stokes equations}

Our new theory starts with the classical Navier-Stokes-Fourier equations which are a differential form of the three classical conservation laws, namely, mass, momentum and energy conservation laws that govern the motion of a fluid. In an Eulerian reference frame they are:

mass balance equation
\begin{equation}
\label{eqn_mass}
\frac{\partial \rho}{\partial t} + \nabla \bcdot [\rho\, U]  = 0,
\end{equation}

momentum balance equation
\begin{equation}
\label{eqn_momentum}
\frac{\partial  \rho \, U }{\partial t}  +  \nabla \bcdot [\rho \,U\otimes\,U]\,+ \,\nabla \bcdot [p \,\pmb{I} \,+ \,\bvarpi^{\rm (NS)}] = 0,
\end{equation}

energy balance equation
\begin{align}
\label{eqn_energy}
 \frac{\partial}{\partial t}  [\frac{1}{2} \rho \,U^2 & + \rho \,e_{\rm{in}}] + \nabla \bcdot [\frac{1}{2} \rho \,U^2\, U + \rho \,e_{\rm{in}} \,U] \nonumber \\
& + \nabla \bcdot [(p \,\pmb{I} + \bvarpi^{\rm (NS)}) \bcdot U] + \nabla \bcdot q^{\rm (NS)} = 0,
\end{align}
where $\rho$ is the mass-density of the fluid, $U$ is the flow mass velocity, $p$ is the hydrostatic pressure, $e_{\rm{in}}$ is the specific internal energy of the fluid, $\bvarpi^{\rm (NS)}$ is the shear stress tensor, $\bi$ is the identity tensor and $q^{\rm (NS)}$ is the heat flux vector. All these hydrodynamic fields are functions of time $t$ and spatial variable $X$. Additionally, $\nabla$ and $\nabla \bcdot$ denote the usual spatial gradient and divergence operators, respectively, while the operator $\otimes$ denotes the usual tensor product of two vectors. Expression for the specific internal energy is given by, $e_{\rm{in}} = p / \rho (\gamma -1)$ with $\gamma$ being the isentropic exponent.
The constitutive models for the shear stress $\bvarpi^{(NS)}$ and the heat flux vector $q^{(NS)}$ are due to Newton's law and Fourier's law, respectively, and they are given by

\begin{align}
\bvarpi^{\rm (NS)} &= -2 \mu \underbrace{\left[\frac{1}{2} (\nabla U + \widetilde{\nabla U}) - \frac{1}{3} \bi \left(\nabla \bcdot U \right) \right]}_{\mathring{\overline{\nabla U}}}, \\
q^{\rm (NS)} &= - \ka \, \nabla T, \label{eqn_stress_hf}
\end{align}
where $\widetilde{\nabla U}$ represents the transpose of $\nabla U$. Coefficients $\mu$ and $\kappa$ are the dynamic viscosity and  the heat conductivity, respectively.

The system \eqref{eqn_mass} - \eqref{eqn_stress_hf} is the well-known conventional fluid flow model and is widely used to model a viscous and heat conducting fluid. Instead of solving directly this system, we first perform a transformation based on the following change of variable:
\begin{align}
\label{eqn_mvel}
U = U_{\rm v} - \ka_{\rm m} \,\nabla \ln \rho  = U_{\rm v} - \frac{\ka_{\rm m}}{\rho} \nabla \rho,
\end{align}
where $\ka_{\rm m}$ is a molecular diffusivity coefficient.

Equation \eqref{eqn_mvel} is a relation between the fluid mass velocity and the fluid volume velocity, $U_{\rm v}$, which originates from \textit{volume diffusion hydrodynamic theory} \cite{Brenner2012,Calgaro2015,DRM2008,Reddyetal2019}. It has also been derived using a stochastic variational method \cite{Koide2018}.

Substituting equation \eqref{eqn_mvel} into the system \eqref{eqn_mass} - \eqref{eqn_stress_hf}, it transforms into a new system which we named the recast Navier-Stokes system and is given by:

recast mass balance equation
\begin{align}
& \frac{\partial \rho}{\partial t}  +  \nabla \bcdot [\rho \, U_{\rm v} - \ka_{\rm m} \,\nabla \rho ]  = 0,  \label{eqn_RNSmass}
\end{align}

recast momentum balance equation
\begin{align}
 \frac{\partial }{\partial t} \left[ \rho \, U_{\rm v} - \ka_{\rm m}\, \nabla \rho \right] & + \nabla \bcdot \left[ \rho \,U_{\rm v} \otimes U_{\rm v} \right] \nonumber \\
& + \nabla \bcdot \left[ p\, \bi + \bvarpi_{\rm v}^{\rm (RNS)} \right] = 0, \label{eqn_RNSmoment}
\end{align}

recast energy balance equation
\begin{align}
& \frac{\partial}{\partial t} \Bigg[\frac{1}{2} \rho\, U_{\rm v}^2 + \rho \, e_{\rm{in}} - \ka_{\rm m}\, \left(\rho\, U_{\rm v} \bcdot \nabla \ln \rho \right) \nonumber \\
& + \frac{1}{2} \ka_{\rm m}^2 \left( \nabla \rho \bcdot \nabla \ln \rho \right) \Bigg] + \nabla \bcdot \left[ \frac{1}{2} \rho \,U_{\rm v}^2 \,U_{\rm v} + \rho\,e_{\rm{in}}\, U_{\rm v} \right] \nonumber \\
& +\,\nabla \bcdot \Big[ \left( p\,\bi + \bvarpi_{\rm v} \right) \bcdot U_{\rm v} - \ka_{\rm m} \,\bvarpi_{\rm v} \bcdot \nabla \ln \rho \Big] \nonumber \\
& + \nabla \bcdot \left[ q^{\rm (RNS)}_{\rm v} + \ka_{\rm m} \,\mcn_{\rm v_1} + \ka_{\rm m}^2\, \mcn_{\rm v_2} + \ka_{\rm m}^3\,\mcn_{\rm v_3} \right] = 0,\label{eqn_RNSenergy}
\end{align}
where the constitutive relations for the new shear stress and the new heat flux vector are given by
\begin{align}
& \bvarpi_{\rm v}^{\rm (RNS)} = \bvarpi_{\rm v} + \frac{\ka_{\rm m}^2}{\rho} \nabla \rho \otimes \nabla \rho - \ka_{\rm m} \,U_{\rm v} \otimes \nabla \rho \nonumber \\
& \qquad \qquad \qquad - \ka_{\rm m} \, \nabla \rho \otimes U_{\rm v}, \label{eqn_RNSstress}\\
& q_{\rm v}^{\rm (RNS)} = q^{\rm (NS)} \,- \,\ka_{\rm m}\, \rho \,e_{\rm{in}}\, \nabla \ln \rho\,-\, \ka_{\rm m}\,p\,\bi \bcdot \nabla \ln \rho, \label{eqn_RNShf}\\
& \textnormal{with} \nonumber \\
& \bvarpi_{\rm v} = - 2 \mu \mathring{\overline{\nabla U_{\rm v}}} \,+\,  2\, \mu \, \ka_{\rm m} \,\widetilde{\bD} \ln \rho\, -\, \frac{2 \mu}{3} \ka_{\rm m} \, \Delta \ln \rho\, \bi, \label{eqn_Tstress}
\end{align}
and $\mcn_{\rm v_i}$ for ${\rm i}=1$ to $3$ represent other nonlinear terms which are given by
\begin{align}
\mcn_{\rm v_1} &= -\,(U_{\rm v} \bcdot \nabla \rho)\, U_{\rm v} \,-\,  \frac{1}{2}\, U_{\rm v}^2 \,\nabla \rho, \\
\mcn_{\rm v_2} &= (U_{\rm v} \bcdot \nabla \rho) \, \nabla \ln \rho \,+\, \frac{1}{2\,\rho}\, |\nabla \rho|^2 \, U_{\rm v}, \\
\mcn_{\rm v_3} &= - \frac{1}{2\, \rho}  |\nabla \rho|^2 \, \nabla \ln \rho.
\end{align}
The operators $\widetilde{\bD}$ and $\Delta$ appearing in \eqref{eqn_Tstress} denote the Hessian and the Laplacian operators, respectively.

The continuum flow system \eqref{eqn_RNSmass} - \eqref{eqn_RNSenergy} is a type of mass diffusion hydrodynamic model.
That is, it contains: (i) a mass diffusion component in the conservation of mass equation, (ii) explicit fluid dilation terms in the momentum stress tensor, and (iii) non-Fourier heat flux terms. It can be converted back into the original system \eqref{eqn_mass} - \eqref{eqn_stress_hf} by reversing the change of variable in equation \eqref{eqn_mvel}. Next, we show that the transformed system \eqref{eqn_RNSmass} - \eqref{eqn_RNSenergy} may be more appropriate to solve for flows involving large density variations/gradients and compare with experimental data.
\section{\label{3}The shock wave structure problem in a monatomic gas}

We consider a planar stationary shock wave propagating in the positive $x$-direction which is established in a flow of a monatomic gas. We denote the upstream ($x \rightarrow -\infty$) and downstream ($x \rightarrow \infty$) conditions of a shock, located at $x = 0$, by a subscript $1$ and $2$, respectively. These upstream and downstream states of the shock are connected by jump conditions: the Rankine-Hugoniot (RH) conditions \cite{CF1948,LR1957}. For this one-dimensional stationary shock flow configuration, the recast Navier-Stokes equations reduced to:
\begin{align}
& \dx \left[ \rho \, u_{\rm v}  - \ka_{\rm m}  \frac{{\rm d} \rho}{{\rm d} x} \right] = 0,\label{eqn_RNSPSW1}\\
& \dx \left[ \rho\, u_{\rm v}^2 + p + \varPi_{\rm v}^{\rm (RNS)}\right] = 0, \label{eqn_RNSPSW2}\\
& \dx \Bigg[ \rho \, u_{\rm v} \,\left(\frac{1}{2}\,u_{\rm v}^2 \,+\,C_p\,T \right)\,\nonumber \\
& \qquad + \,  \left( \varPi_{\rm v} - \frac{3}{2} \ka_{\rm m} \,u_{\rm v}\,\frac{{\rm d} \rho}{{\rm d} x} \right) \, \left(u_{\rm v} - \,\frac{\ka_{\rm m}}{\rho} \frac{{\rm d} \rho}{{\rm d} x} \right) \nonumber \\
& \qquad - \frac{\ka_{\rm m}^3}{2\,\rho^2} \left( \frac{{\rm d} \rho}{{\rm d} x} \right)^3 \,+\,q_{\rm v}^{\rm (RNS)} \Bigg] = 0,\label{eqn_RNSPSW3}
\end{align}
with the only non-zero longitudinal new shear stress $\varPi_{\rm v}^{\rm (RNS)}$ and the new heat flux vector $q_{\rm v}^{\rm (RNS)}$ given by
\begin{align}
& \varPi_{\rm v}^{\rm (RNS)} = \varPi_{\rm v} - 2\,\ka_{\rm m}\,u_{\rm v}\,\frac{{\rm d} \rho}{{\rm d} x} + \frac{\ka_{\rm m}^2}{\rho} \left( \frac{{\rm d} \rho}{{\rm d} x} \right)^2,\label{eqn_RNSPSW4}\\
& \varPi_{\rm v} = -\frac{4}{3} \mu \, \frac{{\rm d} u_{\rm v}}{{\rm d} x} + \frac{4}{3} \frac{\mu\,\ka_{\rm m}}{\rho} \, \frac{{\rm d}^2 \rho}{{\rm d} x^2} - \frac{4}{3} \frac{\mu\,\ka_{\rm m}}{\rho^2} \left( \frac{{\rm d} \rho}{{\rm d} x} \right)^2, \label{eqn_RNSPSW5}\\
& q_{\rm v}^{\rm (RNS)} = -\ka \, \frac{{\rm d} T}{{\rm d} x} - \frac{\gamma}{\left(\gamma - 1\right)} \ka_{\rm m}\, \frac{p}{\rho} \frac{{\rm d} \rho}{{\rm d} x}.\label{eqn_RNSPSW6}
\end{align}
Integration of the system \eqref{eqn_RNSPSW1} - \eqref{eqn_RNSPSW3} and later employing the ideal gas equation of state leads to:
\begin{align}
& \rho \, u_{\rm v} = m_0 + \ka_{\rm m} \,\frac{{\rm d} \rho}{{\rm d} x},\label{eqn_RNSPSW7}\\
& \rho\, R \,T + \rho\,u_{\rm v}^2 +  \varPi_{\rm v}^{\rm (RNS)} = p_0, \label{eqn_RNSPSW8}\\
&  \rho \, u_{\rm v} \left(C_p\, T + \frac{u_{\rm v}^2}{2} \right) \nonumber \\
& + \left( \varPi_{\rm v} - \frac{3}{2} \ka_{\rm m} u_{\rm v} \frac{{\rm d} \rho}{{\rm d} x} \right) \left(u_{\rm v} - \frac{\ka_{\rm m}}{\rho} \frac{{\rm d} \rho}{{\rm d} x} \right) \nonumber \\
& - \frac{\ka_{\rm m}^3}{2\,\rho^2} \left( \frac{{\rm d} \rho}{{\rm d} x} \right)^3 + q_{\rm v}^{\rm (RNS)}  = m_0\,h_0,\label{eqn_RNSPSW9}
\end{align}
where $m_0, p_0$ and $h_0$ are constants which represent the mass flow rate, the stagnation pressure and the stagnation specific enthalpy, respectively. The specific gas constant is denoted by $R$.

In order to solve the system \eqref{eqn_RNSPSW7} - \eqref{eqn_RNSPSW9}, it is convenient to work with its dimensionless form.
We use the following set of dimensionless variables based on the upstream reference states (denoted with subscript $1$):
\begin{align}
\begin{split}
\label{eqn_refvar}
& \overline{\rho} = \frac{c_1^2}{p_1} \rho= \frac{\gamma}{\rho_1}\,\rho, \quad \overline{u}_{\rm v} = \frac{u_{\rm v}}{c_1}, \quad \overline{T} = \frac{R}{c_1^2} T,  \\
& \overline{x} = \frac{x}{\la_1},\quad \overline{\mu} = \frac{\mu}{\mu_1},
\end{split}
\end{align}
where $\lambda_1$ is the upstream mean free path which is a natural choice for a characteristic length-scale as changes through the shock occur due to few collisions. Furthermore, $c_1 = \sqrt{\gamma\, R\, T_1}$ being the adiabatic sound speed.
Further, we assume that the molecular mass diffusivity coefficient $\ka_{\rm m}$ is related to the viscosity coefficient via the relation, $\ka_{\rm m} = \ka_{\rm m_0}\, \mu/\rho$ with $\ka_{\rm m_0}$ being a constant. Hence, the dimensionless forms of transport coefficients $\overline{\ka}$ and $\overline{\ka}_{\rm m}$ are:
\begin{equation}
\overline{\ka} = \frac{\gamma}{(\gamma - 1)\,\rm{Pr}} \overline{\mu} \quad \text{and} \quad \overline{\ka}_{\rm m} =  \ka_{\rm m_0} \frac{\overline{\mu}}{\overline{\rho}},
\end{equation}
where $\rm{Pr}$ is the Prandtl number whose value is equal to $2/3$ for the case of a monatomic gas.

It is well-known that the viscosity and temperature relation has a noticeable effect on the shock wave structure.
Here we adopt the generally accepted temperature-dependent viscosity power law \cite{LC1992,Greenshields2007}: $\mu \propto T^s$ or $\mu = \alpha \, T^s$, where $\alpha$ is a constant of proportionality taken to be $\gamma^s$ and the power $s$ for almost all real gases falling between $0.5 \leq s \leq 1$, with the limiting cases, $s = 0.5$ and $s = 1$ corresponding to theoretical gases, namely, the hard-sphere and Maxwellian gases, respectively. In our simulations we use $s = 0.75$ for a monatomic Argon gas.

The final reduced recast Navier-Stokes system in terms of the dimensionless quantities defined via \eqref{eqn_refvar} is:
\begin{align}
& \overline{\rho} \,\overline{u}_{\rm v} - \left(\frac{\gamma}{\la_0}\right) \overline{\ka}_{\rm m} \frac{{\rm d} \overline{\rho}}{{\rm d} \overline{x}} - \overline{m}_0 = 0,\label{eqn_RNSPSW10}\\
& \overline{\rho} \,\overline{u}_{\rm v}^2 + \overline{\rho} \, \overline{T} + \left(\frac{\gamma}{\la_0}\right) \overline{\varPi}_{\rm v}^{\rm (RNS)}  - \overline{m}_0 \, \overline{p}_0 = 0,  \label{eqn_RNSPSW11}\\
&  \overline{\rho} \, \overline{u}_{\rm v} \left( \frac{\gamma}{\left(\gamma-1\right)}\, \overline{T} + \frac{1}{2} \overline{u}_{\rm v}^2 \right) - \left(\frac{\gamma}{\la_0}\right)^3 \frac{\overline{\ka}_{\rm m}^3}{2\,\overline{\rho}^2} \left( \frac{{\rm d} \overline{\rho}}{{\rm d} \overline{x}} \right)^3 \nonumber \\
& + \left(\frac{\gamma}{\la_0}\right) \left( \overline{\varPi}_{\rm v} - \frac{3}{2} \overline{\ka}_{\rm m}\, \overline{u}_{\rm v} \frac{{\rm d} \overline{\rho}}{{\rm d} \overline{x}} \right) \left(\overline{u}_{\rm v} - \frac{\gamma}{\la_0} \frac{\overline{\ka}_{\rm m}}{\overline{\rho}} \frac{{\rm d} \overline{\rho}}{{\rm d} \overline{x}} \right) \nonumber \\
& + \left(\frac{\gamma}{\la_0}\right) q_{\rm v}^{\rm (RNS)}  - \overline{m}_0\,\overline{h}_0 = 0.\label{eqn_RNSPSW12}
\end{align}
The upstream Mach number is defined as, ${\it{M}}_{\rm 1} = u_{\rm v_1}/c_1$. Quantities $\overline{m}_0$, $\overline{p}_0$ and $\overline{h}_0$ are  integration constants whose expressions are obtained using the well-known Rankine-Hugoniot conditions:
\begin{align}
& \overline{m}_0 = \gamma \, {\it{M}}_{\rm 1}, \\
& \overline{p}_0 = \frac{1}{\gamma\, {\it{M}}_{\rm 1}} \, \left(1\,+\, \gamma\, {\it{M}}_{\rm 1}^{\rm 2} \right),\\
& \overline{h}_0 = 1\,+\,\frac{\left(\gamma\,-1 \right)}{2} {\it{M}}_{\rm 1}^{\rm 2}.
\end{align}
Expressions for the dimensionless new shear stress and the new heat flux are given by
\begin{align}
& \overline{\varPi}_{\rm v}^{\rm (RNS)} = \overline{\varPi}_{\rm v} - 2\, \overline{\ka}_{\rm m}\,\overline{u}_{\rm v} \frac{{\rm d} \overline{\rho}}{{\rm d} \overline{x}} + \left( \frac{\gamma}{\la_0}\right)  \frac{\overline{\ka}_{\rm m}^2}{\overline{\rho}} \left( \frac{{\rm d} \overline{\rho}}{{\rm d} \overline{x}} \right)^2,\label{eqn_RNSPSW15}\\
& \overline{q}_{\rm v}^{\rm (RNS)} = -\overline{\ka} \, \frac{{\rm d} \overline{T}}{{\rm d} \overline{x}} \,-\,\frac{\gamma}{\left(\gamma - 1\right)}\,  \overline{\ka}_{\rm m}\, \overline{T}\, \frac{{\rm d} \overline{\rho}}{{\rm d} \overline{x}},\label{eqn_RNSPSW16}\\
& \text{with} \nonumber \\
& \overline{\varPi}_{\rm v} = -\frac{4}{3} \overline{\mu} \frac{{\rm d} \overline{u}_{\rm v}}{{\rm d} \overline{x}} + \frac{4}{3} \left( \frac{\gamma}{\la_0}\right) \frac{\overline{\mu} \,\overline{\ka}_{\rm m}}{\overline{\rho}} \frac{{\rm d}^2 \overline{\rho}}{{\rm d} \overline{x}^2} \nonumber \\
& \qquad \quad - \frac{4}{3} \left(\frac{\gamma}{\la_0}\right) \frac{\overline{\mu}\, \overline{\ka}_{\rm m}}{\overline{\rho}^2} \left( \frac{{\rm d} \overline{\rho}}{{\rm d} \overline{x}} \right)^2. \label{eqn_RNSPSW14}
\end{align}
We solved the final system \eqref{eqn_RNSPSW10}--\eqref{eqn_RNSPSW12} using a numerical scheme, namely, the finite difference global solution (FDGS) developed by Reese et al. \cite{Reese1995} with well-posed boundary conditions. The specific details of FDGS scheme can be found in \cite{Reese1995}.

\section{\label{4} Results and discussion}
We perform numerical simulations of stationary shock waves located at $x=0$ using the FDGS scheme by considering a computational spatial domain of length $40 \lambda_1$ covering $(-20 \lambda_1, 20 \lambda_1)$ with 275 spatial grid points (for which the spatial convergence is reached). This is wide enough to contain the entire shock profile for $1.55 \leq {\it{M}}_{\rm 1} \leq 9$ without altering its structure. We observed that recast Navier-Stokes computations show numerical oscillations at upstream and downstream parts for certain values of $\ka_{\rm m_0}$ at Mach numbers larger than 6. Therefore, we assume the molecular mass diffusivity coefficient, $\ka_{\rm m}$, to depend on the Mach number. An initial base value for $\ka_{\rm m_0}$ is identified as $\ka_0 = \gamma/\left((\gamma - 1)\,\rm{Pr}\right)$, then the different values used based on it at the various  Mach numbers in  our present results are given in table~\ref{tab:1}.
To compare the shock structure profiles among the theoretical and experimental data, the position $x$ has been scaled such that $x = 0$ corresponds to a value of the normalized gas density, $\rho_{\rm N} = (\rho - \rho_1)/(\rho_2 - \rho_1)$, of $0.5$.

\begin{figure*}
  \includegraphics[width=0.325\textwidth]{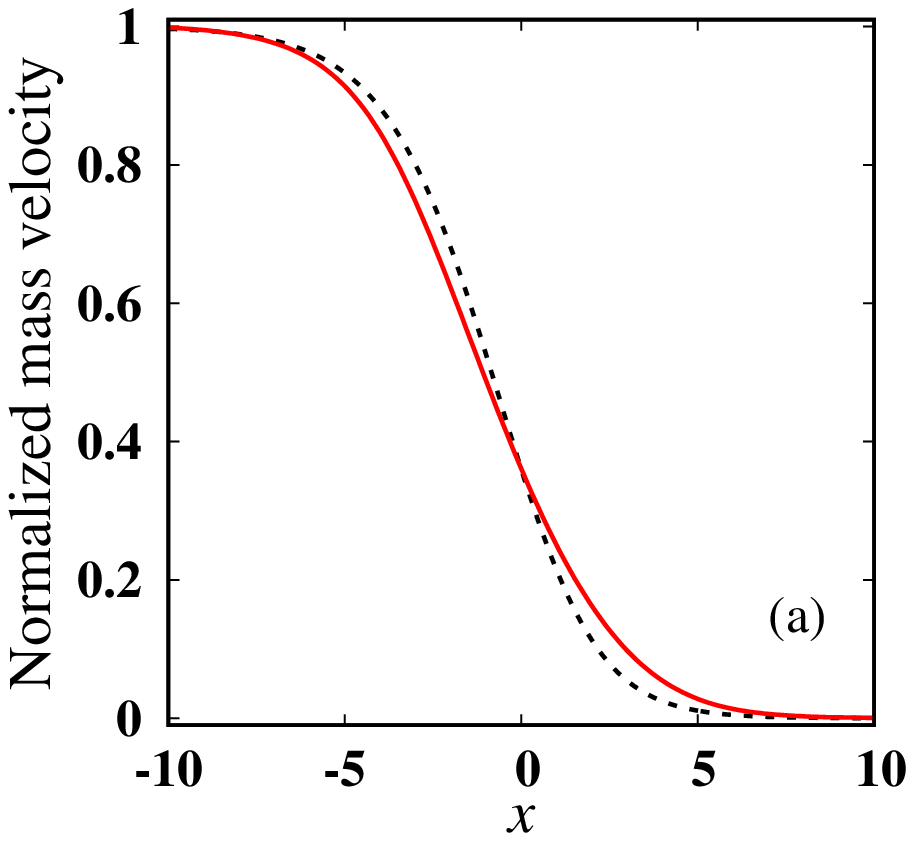}
  \includegraphics[width=0.325\textwidth]{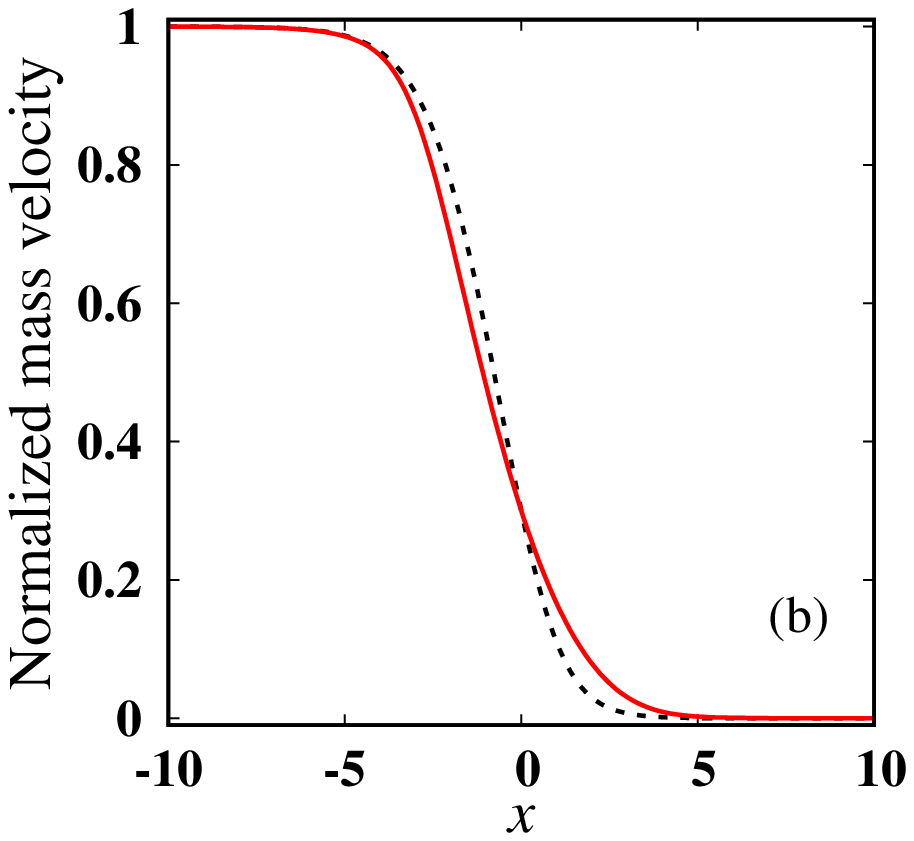}
  \includegraphics[width=0.325\textwidth]{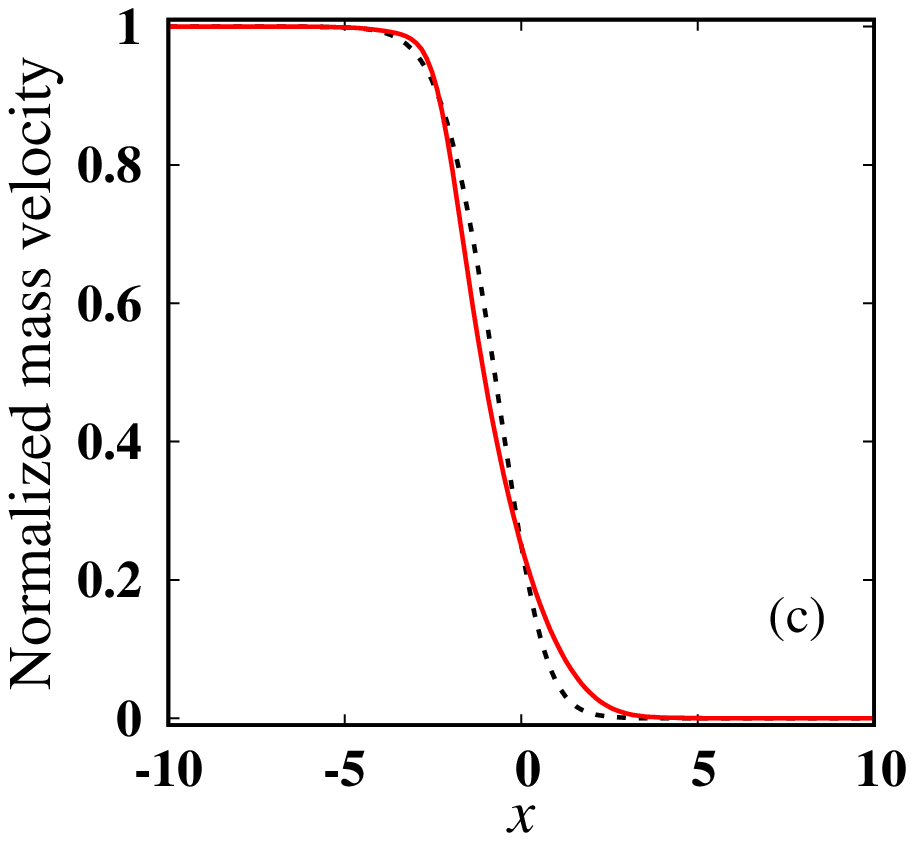}\\
  \includegraphics[width=0.325\textwidth]{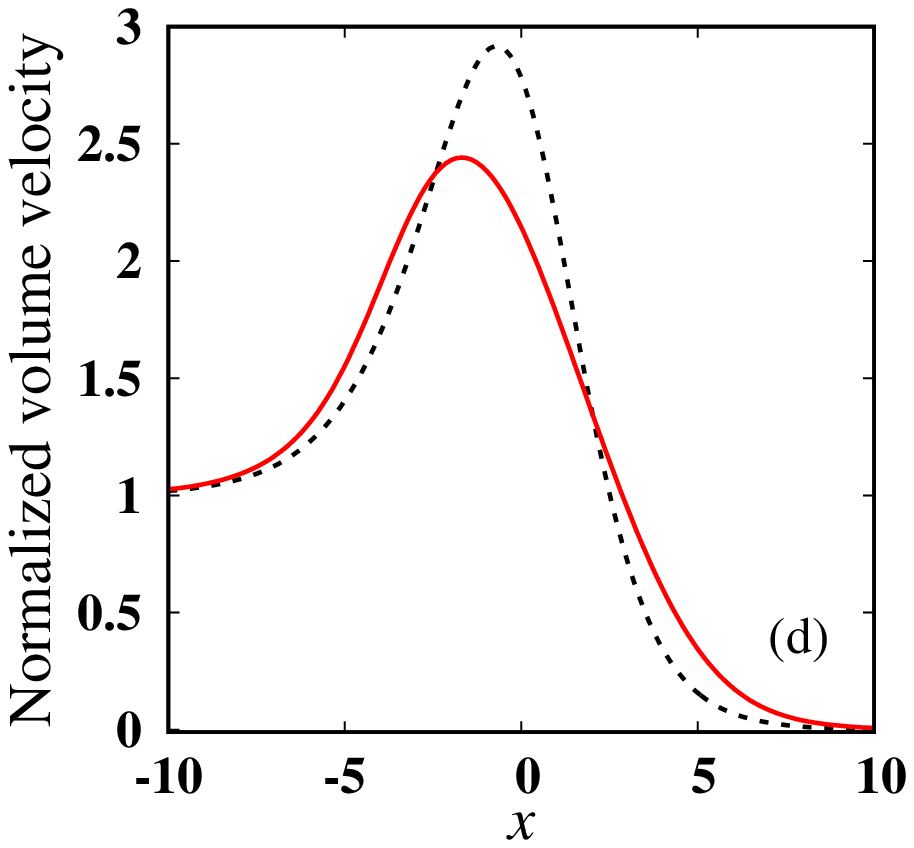}
  \includegraphics[width=0.325\textwidth]{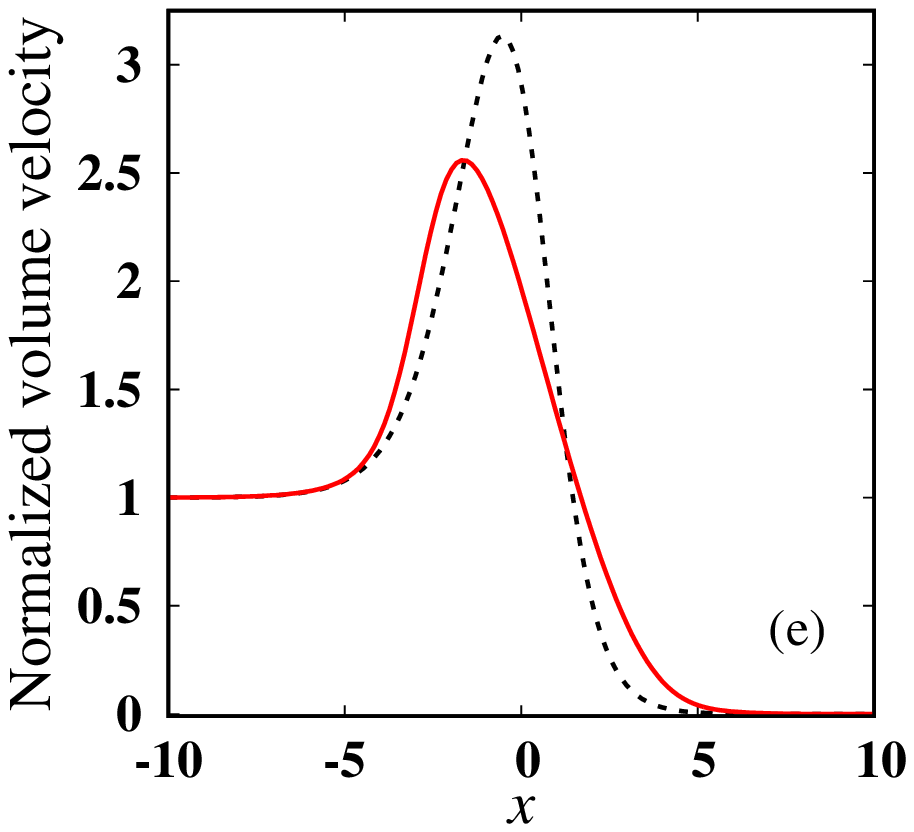}
  \includegraphics[width=0.325\textwidth]{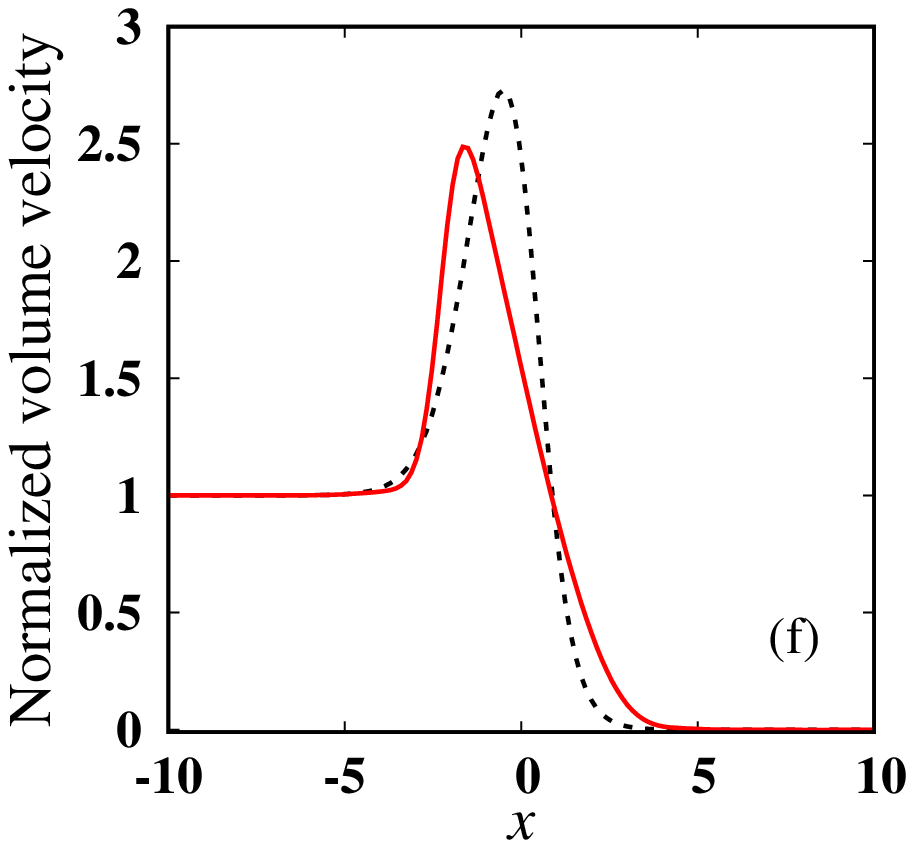}
\caption{Comparison of normalized mass velocity (upper panels) and volume velocity (lower panels) profiles in $\rm{Ar}$ shock layer: for (a,d) ${\it{M}}_{\rm 1} = 1.55$, (b,e) ${\it{M}}_{\rm 1} = 2.05$ and (c,f) ${\it{M}}_{\rm 1} = 3$. In each panel, dashed black line and solid red line indicates the solutions from classical Navier-Stokes and recast Navier-Stokes, respectively.}
\label{fig:1}
\end{figure*}

\begin{table}
\caption{Upstream Mach number range and corresponding value of $\ka_{\rm m_0}$ used in present results.}
\label{tab:1}       
\begin{tabular}{lll}
\hline\noalign{\smallskip}
${\it{M}}_{\rm 1}$ range &  $\ka_{\rm m_0}$ value  \\
\noalign{\smallskip}\hline\noalign{\smallskip}
$1 < {\it{M}}_{\rm 1} < 2$ &  $8 \,\ka_0$  \\
$2\leq {\it{M}}_{\rm 1} < 3$ &  $6 \,\ka_0$  \\
$3 \leq {\it{M}}_{\rm 1} \leq 4$ &  $4 \,\ka_0$  \\
$4 < {\it{M}}_{\rm 1} \leq 5$ &  $3.25 \, \ka_0$  \\
$5 < {\it{M}}_{\rm 1} \leq 6$ &  $3 \,\ka_0$  \\
$6 < {\it{M}}_{\rm 1} \leq 7$ &  $2.75 \, \ka_0$  \\
$ {\it{M}}_{\rm 1} > 7$ &  $2 \, \ka_0$  \\
\noalign{\smallskip}\hline
\end{tabular}
\end{table}

\subsection{\label{4.0} Mass velocity vs volume velocity}
While the recast Navier-Stokes and the original equations may convert into one another, the velocity profile solution from the original represents the mass velocity ($u$), and the transformed equations gives the volume velocity ($u_{\rm v}$). The two differ by the diffusive flux as defined in \eqref{eqn_mvel}. However, one can compute the mass velocity from the recast Navier-Stokes solution and the volume velocity from the classical Navier-Stokes solution using relation \eqref{eqn_mvel}. Panels (a), (b) and (c) of Fig. \ref{fig:1} show the mass velocities predicted by classical and recast Navier-Stokes for upstream Mach numbers of ${\it{M}}_{\rm 1} =  1.55, 2.05$ and $3$, respectively, while panels (d), (e) and (f) of Fig. \ref{fig:1} show the volume velocities for the same upstream Mach numbers. Both velocity profiles have been normalized such that $u_{\rm N} = (u - u_2)/(u_1 - u_2)$ and $u_{\rm v_N} = (u_{\rm v} - u_{\rm v_2})/(u_{\rm v_1} - u_{\rm v_2})$. It is evident from Fig. \ref{fig:1} (a-c) that the mass velocity predicted by the classical and recast Navier-Stokes are not the same at all Mach numbers. The profile of the recast model  mass velocity is flatter than the classical prediction at low Mach numbers and steepens at the upstream part at large Mach numbers (${\it{M}}_{\rm 1} > 3$).
It is seen from Fig. \ref{fig:1} (d-f) that volume velocities from both classical and recast Navier-Stokes overshoot within the shock layer. This is due to the large density gradient involved and the overshoot increases with increasing ${\it{M}}_{\rm 1}$. This overshoot shows that the change of variables expressed by relation \eqref{eqn_mvel} is not smooth. Next we show that not only the velocity profiles differ in the transformation process but the entire hydrodynamic field variables compare differently with experiments.

\subsection{\label{4.1} Density profiles}
Full experimental data exist for monatomic gas density variations within shock layers \cite{Alsmeyer1976}. These data are therefore our first choice for comparison. They are obtained for shock waves in Argon for upstream Mach numbers ranging from $1.55$ to $9$.

\begin{figure*}
  \includegraphics[width=0.475\textwidth]{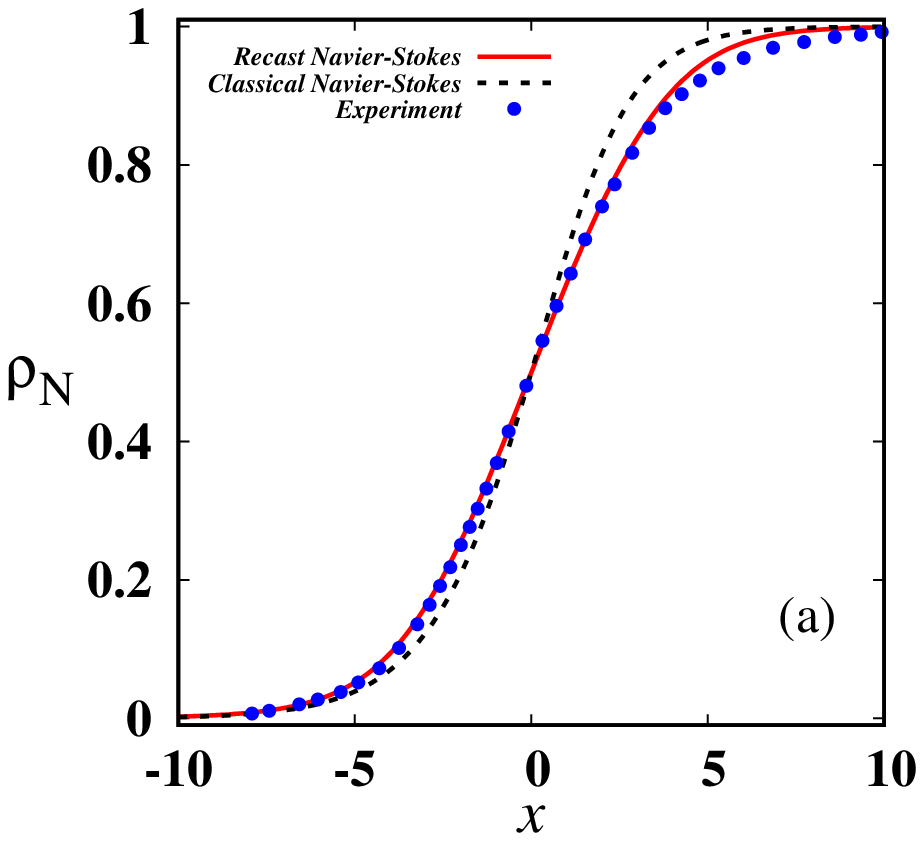}
  \includegraphics[width=0.475\textwidth]{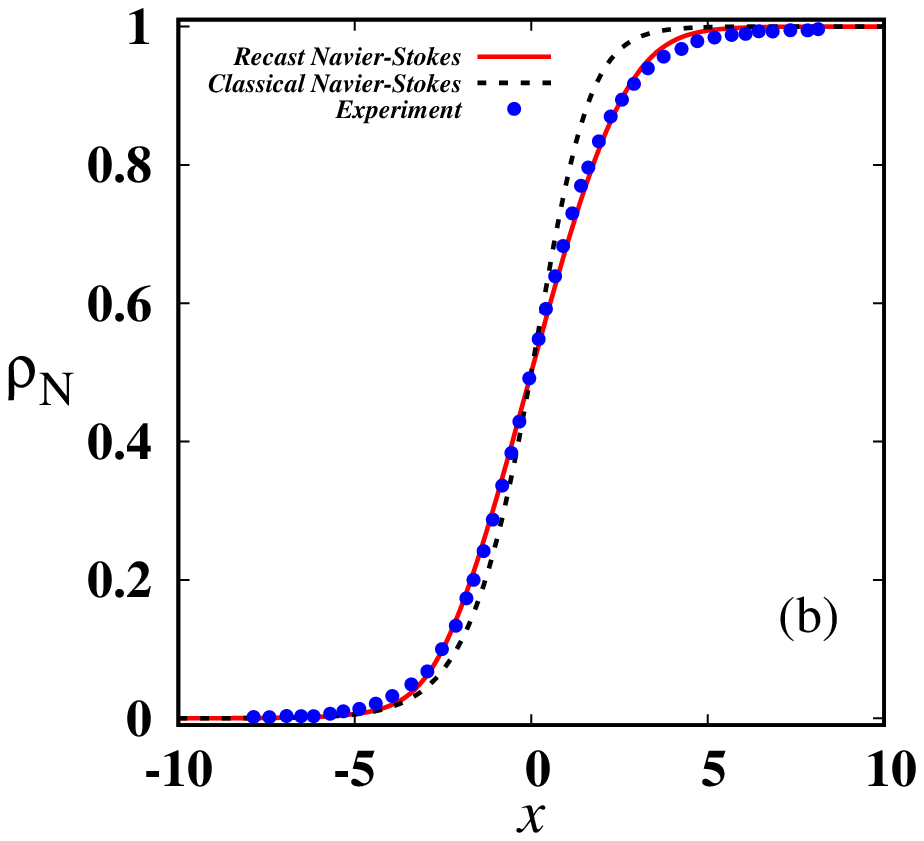}
  \includegraphics[width=0.475\textwidth]{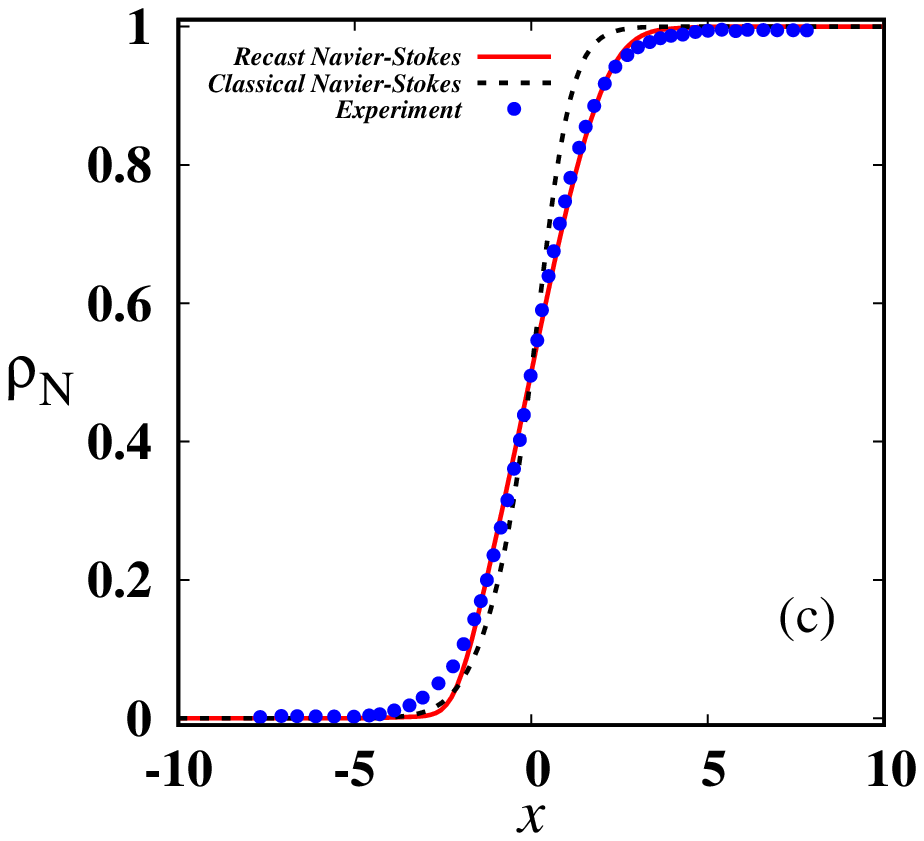}
  \includegraphics[width=0.475\textwidth]{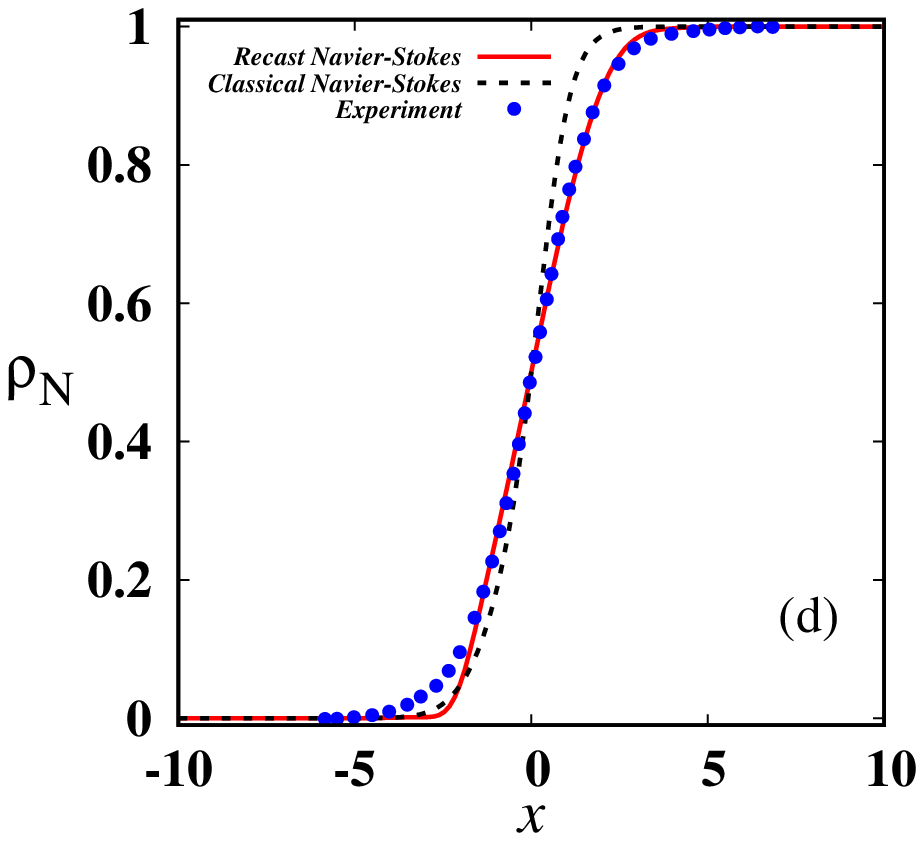}
  \includegraphics[width=0.475\textwidth]{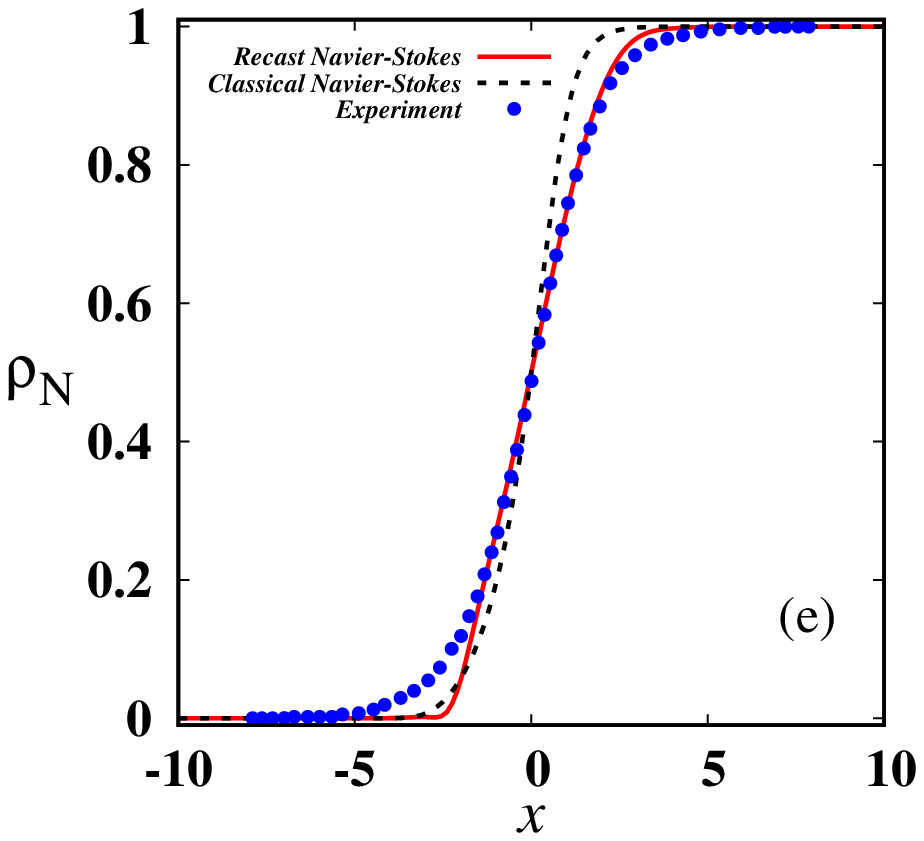}\qquad \,
  \includegraphics[width=0.475\textwidth]{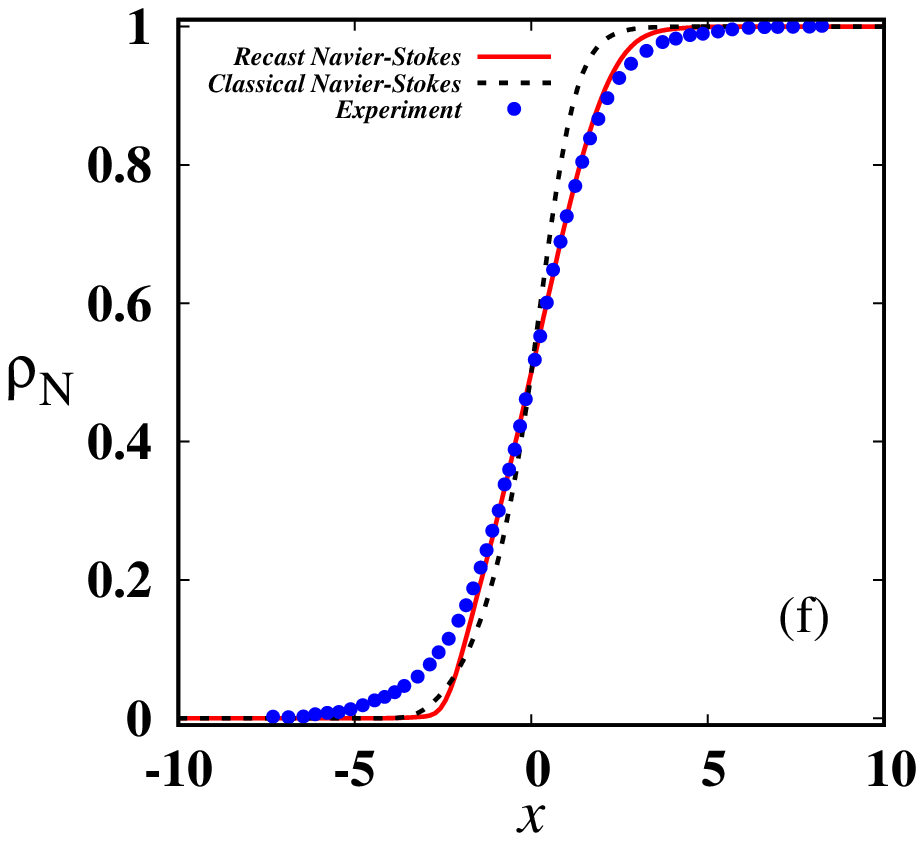}
\caption{Variation of normalized density ($\rho_N$) profiles in $\rm{Ar}$ shock layer: for (a) ${\it{M}}_{\rm 1} = 1.55$, (b) ${\it{M}}_{\rm 1} = 2.05$, (c) ${\it{M}}_{\rm 1} = 3.38$, (d) ${\it{M}}_{\rm 1} = 3.8$, (e) ${\it{M}}_{\rm 1} = 6.5$ and (f) ${\it{M}}_{\rm 1} = 9$. In each panel, dashed black line represents the solution of the classical Navier-Stokes equations, solid red line represents the solutions of the recast Navier-Stokes equations and filled blue circles represent experimental data of Alsmeyer \cite{Alsmeyer1976}.}
\label{fig:2}
\end{figure*}

Figure \ref{fig:2} shows the predicted normalized density profiles through an Argon shock wave using the recast and the original equations with $s = 0.75$ compared with the experimentally measured density data. Panels (a), (b), (c), (d), (e) and (f) of Fig. \ref{fig:2} correspond to upstream Mach numbers of ${\it{M}}_{\rm 1} = 1.55, 2.05, 3.38$, $3.8, 6.5,$ and $9$, respectively. In each panel, the dotted black lines represent solutions of the Navier-Stokes equations and the solid red lines represent solutions by the recast Navier-Stokes equations. The filled blue circles represent the experimental data. For the upstream Mach number of $1.55$, one observes from panel (a) of Fig. \ref{fig:2} that the classical Navier-Stokes equations almost predict the upstream shock layer as the experiments but completely fail to predict the downstream shock layer. The recast Navier-Stokes equations produce very good agreement with the experimental data with a small disparity at the downstream shock layer.
The recast Navier-Stokes predictions for the normalized density profiles show excellent agreement with the experimental data for the upstream Mach number of ${\it{M}}_{\rm 1} = 1.55, 2.05$, and $3.38$, which is evident from panels (a)-(c) of Fig. \ref{fig:2}. In fact, a good agreement between predictions of the recast Navier-Stokes equations and the experimental data of Alsmeyer \cite{Alsmeyer1976} is found for upstream Mach numbers up to about $3.8$. At the high upstream Mach numbers ${\it{M}}_{\rm 1} = 6.5$, Fig. \ref{fig:2}(e) and ${\it{M}}_{\rm 1} = 9$, Fig. \ref{fig:2}(f), the predictions of recast Navier-Stokes equations for the variation of the density within the shock layer are still better compared to the original equations. At the upstream Mach number of $6.5$ and $9$, the recast Navier-Stokes predictions are not as flat as the experimental predictions at the upstream part of the shock but with visible excellent match downstream. Overall, the recast Navier-Stokes solutions show better agreement with experimental values than the original at all upstream Mach numbers.

\subsection{\label{4.2} Reciprocal shock thickness}
Generally, studies of shock structures include a validation by comparing a few shock structure parameters with experimental data, where available, and  other numerical simulations. One of the principal parameters of shock structure is the non-dimensional inverse shock thickness $\delta = \la_1/L$, where the shock thickness or shock width is defined as \cite{Greenshields2007,Alsmeyer1976}:
\begin{equation}
\label{eqn_SW}
L = \frac{\rho_2 - \rho_1}{|\max(\frac{{\rm d} \rho}{{\rm d} x})|}.
\end{equation}
This definition is based on the density profile and depends mainly on the central part of the shock wave.
The reciprocal shock thickness ($\delta$) is one of the widely used shock parameters to compare computational results with experiments as it possesses an important feature that is, it actually represents the Knudsen number of the shock structure flow problem. In other words, the shock thickness acts as the characteristic dimension of the flow configuration \cite{Greenshields2007}.

\begin{figure*}
  \includegraphics[width=0.325\textwidth]{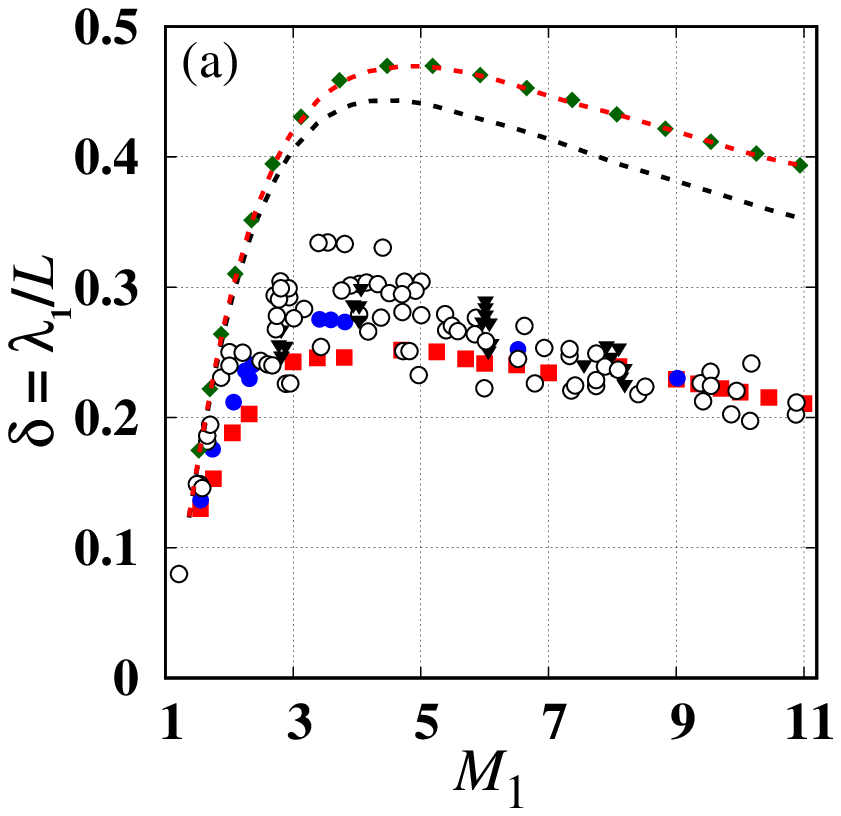}
  \includegraphics[width=0.325\textwidth]{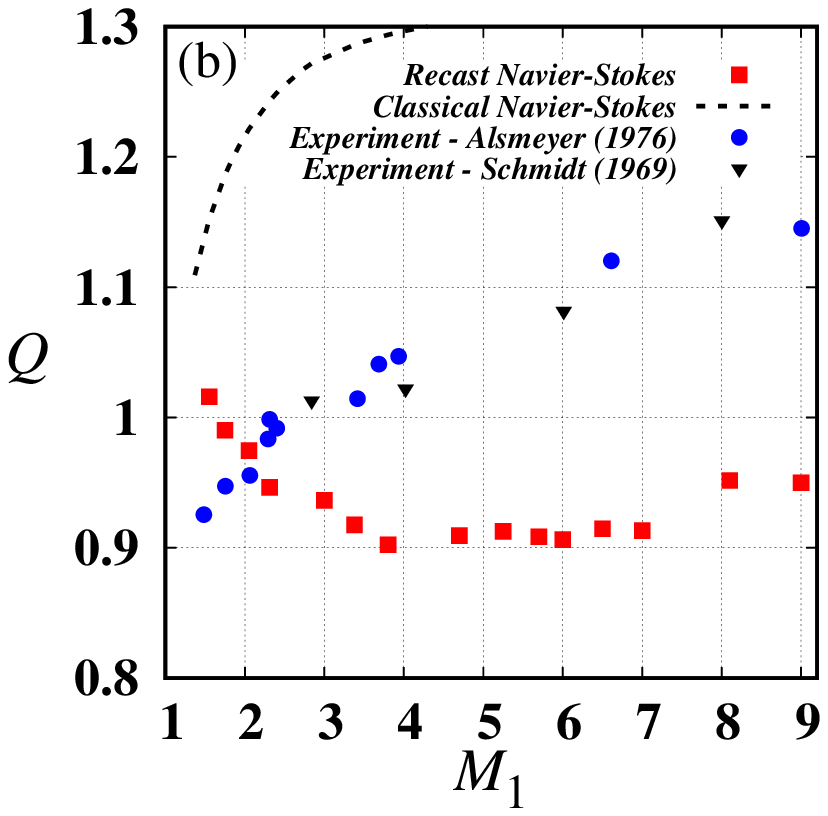}
  \includegraphics[width=0.325\textwidth]{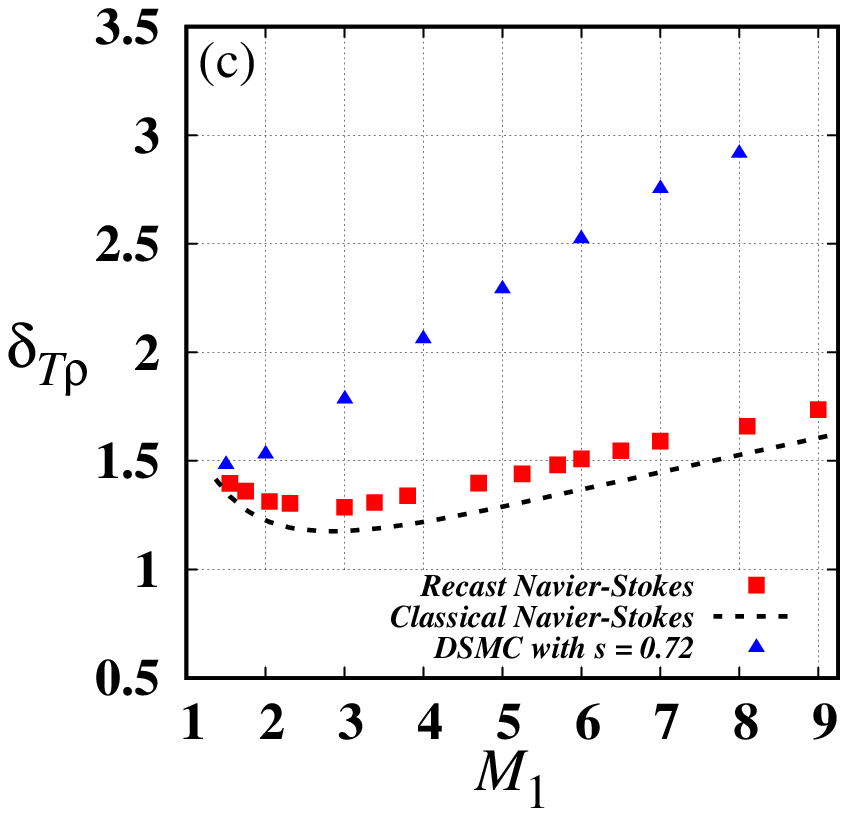}
\caption{Variation of the different shock structure parameters for monatomic gas, $\rm{Ar}$: (a) reciprocal shock width $(\delta)$, (b) density asymmetry quotient $(Q)$, and (c) temperature-density spatial lag $(\delta_{T\rho})$. Theoretical results: $\color{red}\blacksquare$ -- recast Navier-Stokes solution with $\ka_{\rm m_0}$ as in table $1$ and $s = 0.75$, black and red dashed lines show present solutions of NS with $s = 0.75$ and $s = 0.72$, respectively, using FDGS technique, $\color{darkgreen}\blacklozenge$ -- NS solution from \cite{Paolucci2018} with $s = 0.72$; experimental results: $\color{blue}\bullet$ -- Alsmeyer \cite{Alsmeyer1976}, $\blacktriangledown$ -- Schmidt \cite{Schmidt1969} and {\color{black}\textopenbullet} -- other experimental data assembled from \cite{Alsmeyer1976}; $\color{blue}\blacktriangle$ -- DSMC results with $s = 0.72$ taken from \cite{LC1992}.}
\label{fig:3}
\end{figure*}

The most comprehensive collection of experimental data for the reciprocal shock thickness ($\delta$) is reported in \cite{Alsmeyer1976}. Figure \ref{fig:3}(a) shows predictions of recast Navier-Stokes equations for the reciprocal shock thickness (the inverse density thickness) in Argon for an upstream Mach number up to ${\it{M}}_{\rm 1} = 11$, with experimental data assembled from \cite{Alsmeyer1976}. Predictions from the classical Navier-Stokes are also presented for the sake of completeness. It is seen in Fig. \ref{fig:3}(a) that our numerical result on the reciprocal shock thickness using classical Navier-Stokes with $s = 0.72$ (red dotted line) coincides with the result from \cite{Paolucci2018} (green rhombus symbols). This confirms the accuracy of the current numerical scheme (FDGS technique). From Fig. \ref{fig:3}(a), one can observe that the classical Navier-Stokes equations with $s = 0.75$ (black dotted line) and with $s = 0.72$ (red dotted line) predict the reciprocal shock thickness to be $1.4$ to $2$ times the measured value over the entire Mach number range presented. However, the solution from the recast Navier-Stokes equations with the choice of $\ka_{\rm m_0}$ values listed in table \ref{tab:1} and $s = 0.75$ is found to follow the experimental results of \cite{Alsmeyer1976}. It is noteworthy to mention that for $\ka_{\rm m_0} = 0$, results using the recast NS coincide with that of the classical NS.

\subsection{\label{4.3} Asymmetry quotient of density profile}

From Fig. \ref{fig:2}, at the upstream and downstream part of the profile one can observe that there are still some discrepancies between predictions and experimental shock density profiles. 
However, the results by the reciprocal shock thickness $\delta$ conclude that the recast Navier-Stokes equations show excellent agreement with the experimental data. This suggests that the inverse density thickness $\delta$ does not express full information about the overall shape of the shock wave profile, as it just depends on the maximum density gradient alone.

A second important measure of a shock structure for which experimental results are available is the asymmetry of the density profile, $Q$. This gives more information about the shape of the shock profile as it measures skewness of the density profile relative to its midpoint \cite{Greenshields2007}. The shock asymmetry, $Q$, is defined based on the normalized density profile, $\rho_N$, with its centre, $\rho_N = 0.5$, located at $x = 0$, as
\begin{equation}
\label{eqn_Asym}
Q = \frac{\int_{-\infty}^{0} \rho_{N}(x)\, dx}{\int_{0}^{\infty} \left[ 1\,-\,\rho_N(x)\right] dx}.
\end{equation}
From definition \eqref{eqn_Asym} it is clear that a symmetric shock wave will have a density asymmetry quotient of unity, while for realistic shock waves its value is around unity as shocks are not completely symmetric about their midpoint. Figure \ref{fig:3}(b) shows predictions of the recast Navier-Stokes and the Navier-Stokes equations for the asymmetry quotient compared with experimental data of Alsmeyer \cite{Alsmeyer1976} and Schmidt \cite{Schmidt1969}.
The classical Navier-Stokes equations predict an asymmetry quotient of more than unity at all Mach numbers and these results are not at all in agreement with the experiments. This is evident from panel (b) of Fig. \ref{fig:3}. The recast Navier-Stokes predict an asymmetry quotient of around unity with less than 10 \% deviation from unity at all upstream Mach numbers $(0.9\lessapprox Q \lessapprox 1)$. From this, one can conclude that density profiles predicted by the recast Navier-Stokes are almost symmetric about their midpoint.

\subsection{\label{4.4} Spatial lag of temperature-density profiles}
Another shock structure parameter is defined based on the spatial difference between the temperature and density shock profiles.
Due to the different finite relaxation times between momentum transport and energy transport, variation in density and temperature within a shock does not occur at the same time. Spatial density changes occur after temperature changes. Hence, the spatial difference, $\delta_{T\rho}$, between the normalized density and temperature profiles is defined by
\begin{equation}
\label{eqn_Trhosep}
\delta_{T\rho} = \arrowvert x(0.5\, T_{\rm N}) - x(0.5\, \rho_{\rm N}) \arrowvert,
\end{equation}
where $T_{\rm N} = (T - T_1)/(T_2 - T_1)$ is the normalized temperature. From definition \eqref{eqn_Trhosep} it is clear that the temperature - density separation measures the distance between the midpoints of the respective normalized profiles. Due to lack of experimental data for this shock structure parameter, we utilize available DSMC data \cite{Greenshields2007,LC1992} to compare with the predictions by the theoretical models.

Figure \ref{fig:3}(c) compares results between the recast and the classical Navier-Stokes equations along with DSMC data of Lumpkin and Chapman \cite{LC1992} for the shock macroscopic parameter temperature-density separation, $\delta_{T\rho}$. From panel (c) of Fig. \ref{fig:3} it can be seen that the DSMC data with a viscosity-temperature exponent $s = 0.72$ show that the $\delta_{T\rho}$ value increases with increasing Mach number, in particular, it increases from $\approx 1.5$ to $\approx 2.9$ when the Mach number increases from $1.5$ to $8$.
Results obtained with recast Navier-Stokes equations quantitatively follow that of the classical equations. Both classical and recast Navier-Stokes equations under-predict $\delta_{T\rho}$ at all upstream Mach numbers. One can observe that the hydrodynamic equations (classical and recast) show a decreasing $\delta_{T\rho}$ for $1.5\leq {\it{M}}_{\rm 1} \leq 3$, and then the value of $\delta_{T\rho}$ increases for ${\it{M}}_{\rm 1} > 3$. Generally, as explicit experimental data are not available for temperature profiles it is inconvenient to conclude which model predicts the accurate temperature-density separation from Fig. \ref{fig:3}(c).

\section{\label{5} Conclusions}
The stationary shock structure problem in a monatomic gas (Argon) is analyzed by numerically solving the classical and recast Navier-Stokes equations. We observed that solutions as given by the recast Navier-Stokes equations differ from the solutions by the original equations. The difference is attributable to the fact that hydrodynamic field variables from the recast equations no longer operate as in the original equations (also as boundary conditions are set based on redefined hydrodynamic variables rather than those in the original equations, see ref. \cite{Stamatiouetal2019}). The recast Navier-Stokes equations with a Mach number-dependent mass diffusion coefficient, $\ka_{\rm m_0}$ (see table \ref{tab:1} for its values), and a viscosity-temperature exponent, $s = 0.75$, show better agreements with Alsmeyer's \cite{Alsmeyer1976} experimentally measured density profiles in Argon gas. In the case of the reciprocal shock thickness, the recast Navier-Stokes delivered a good match with the experimental data, and the results exactly coincide with the experimental data at large upstream Mach numbers. However, it does not reproduce the more detailed density asymmetry quotient and temperature-density separation. Nevertheless, we conclude that the recast Navier-Stokes equations better reproduce the shock profiles experimental data. We therefore suggest further investigation and examination of the recast model on other non-equilibrium gas flow configurations.
\begin{acknowledgements}
This research is supported by the UK's Engineering and Physical Sciences Research Council (EPSRC) under grant no. EP/R008027/1 and The Leverhulme Trust, UK, under grant Ref. RPG-2018-174. The authors also thank Jonathan Betts for contributing to checking the simulations.
\end{acknowledgements}

%
%



\end{document}